**ARTICLE TYPE**

# Stencil Computations on AMD and Nvidia Graphics Processors: Performance and Tuning Strategies

Johannes Pekkilä[1] | Oskar Lappi[2] | Fredrik Robertsén[3] | Maarit J. Korpi-Lagg[1,4,5]

[1]Department of Computer Science, Aalto University, Espoo, 02150, Finland

[2]Department of Computer Science, University of Helsinki, Helsinki, 00560, Finland

[3]CSC – IT Center for Science Ltd., Espoo, 02150, Finland

[4]Max Planck Institute for Solar System Research, Göttingen, D-37077, Germany

[5]Nordita, KTH Royal Institute of Technology and Stockholm University, Stockholm, SE-10691, Sweden

**Correspondence**

Corresponding author Johannes Pekkilä, Department of Computer Science, Aalto University, Espoo, 02150, Finland.
Email: johannes.pekkila@aalto.fi

**Abstract**

Over the last ten years, graphics processors have become the de facto accelerator for data-parallel tasks in various branches of high-performance computing, including machine learning and computational sciences. However, with the recent introduction of AMD-manufactured graphics processors to the world's fastest supercomputers, tuning strategies established for previous hardware generations must be re-evaluated. In this study, we evaluate the performance and energy efficiency of stencil computations on modern datacenter graphics processors, and propose a tuning strategy for fusing cache-heavy stencil kernels. The studied cases comprise both synthetic and practical applications, which involve the evaluation of linear and nonlinear stencil functions in one to three dimensions. Our experiments reveal that AMD and Nvidia graphics processors exhibit key differences in both hardware and software, necessitating platform-specific tuning to reach their full computational potential.

## 1 | INTRODUCTION

Stencil computations belong to a class of algorithms, where the elements of an array are updated by extracting information from their neighborhood in a fixed pattern, called a stencil (Fig. 1). A typical example is median filtering in image processing, where the color of each pixel is set to the median color of its neighbors. Other examples include cellular automata[1] and direct numerical simulations based on finite differences[2]. Stencil computations also form the core of convolutional neural networks (CNNs)[3,4], where the graphical representation of the non-zero values of a convolution kernel is a stencil. Efficient stencil computation is desirable, because it enables, for instance, higher resolution in fluid simulations and reduced training time of CNNs.

The introduction of graphics processing units (GPUs) to high-performance computing (HPC) has enabled significantly higher throughput in data-parallel workloads, including stencil computations[5,6]. In comparison to central processing units (CPUs), GPUs are capable of providing higher operational performance and memory bandwidth due to their specialized architecture designed originally for accelerating computations in real-time computer graphics, where a typical load consists of applying a linear transformation on an array of vectors, i.e. a projection matrix on the vertices of a polygon mesh or a filter kernel on a two-dimensional image. In addition to computer graphics, GPUs have been deployed successfully for accelerating computational physics[7,8,9,10], chemistry[11,12], and machine learning[13,14]. The increase in computational power provided by massively parallel microprocessors is considered one of the key factors in the success of deep neural networks over the last decade[15].

A notable feature of the hardware is that it provides a tremendous amount of compute performance in comparison to memory bandwidth. For example, to saturate both the compute and memory units of an Nvidia A100 GPU, a program must execute 50 FP64 operations for each 8-byte

**Abbreviations:** CNN, convolutional neural network; CPU, central processing unit; CU, compute unit; DSL, domain-specific language GCD, graphics compute die; GPU, graphics processing unit; HPC, high-performance computing; HWC, hardware-managed caching; ILP, instruction-level parallelism; LDS, local data share; PDE, partial differential equation; SWC, software-managed caching; TLP, thread-level parallelism.

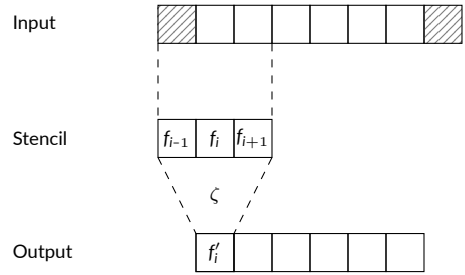

**FIGURE 1** Illustration of stencil computations. A cell represents an array element. The hatched cells represent the padding around the computational domain needed to preserve the shape of the input array across iterations. The values of the padded elements depend on the requirements of the application, for example, boundary conditions when computing PDEs. Here $\zeta$ is a stencil function that maps three elements centered around the point of interest $i$ to an output $f'_i$. We use the prime symbol to denote the updated state throughout this work. The computations progress by sliding the stencil across the input array until all output elements have been mapped.

word transferred from off-chip memory [16]. If the computations require fewer operations per transfer, a significant portion of performance tuning will revolve around maximizing the efficiency of off-chip memory transfers by blocking data in the registers and user-programmable cache.

In this study, we explore the performance, energy efficiency, and tuning features of the A100, MI100, MI250X, and V100 graphics processors in one- to three-dimensional stencil computations. We make the following contributions.

- Measure the performance and energy efficiency of the A100, MI100, MI250X, and V100 graphics processors in linear and nonlinear stencil computations, placing special focus on the performance under heavy cache loads.
- Investigate how the architectural differences of the A100, MI100, MI250X, and V100 graphics processors affect the effectiveness of common tuning strategies. We evaluate the use of hardware- and software-managed caches as the primary methods for data reuse.
- Explore the use of data tensors, convolutions, and other computational tools typically featured in CNNs for computing systems of partial differential equations.
- Propose enhancements to the kernel fusion technique that increase instruction-level parallelism and on-chip data reuse in cache-heavy nonlinear stencil computations.

The remainder of this article is organized as follows. In Section 2, we give an overview of the architecture of modern graphics processors and introduce stencil computations. In Section 3, we outline the methodology and describe the test problems used for evaluating the hardware architectures. The technical implementations of the test problems are presented in Section 4. The results are presented and disseminated in Sections 5 and 6. Section 7 concludes the article.

## 2 | BACKGROUND

In this section, we give an introduction to GPU architecture and stencil computations required for understanding the remainder of this article.

## 2.1 | Terminology

Throughout this work, we use the terms *thread*, *warp*, *thread block*, and *shared memory*, associated with AMD's HIP and Nvidia's CUDA programming models [17,18]. We use the terms *compute unit* and *device* based on Khronos Group's OpenCL specification to refer to the conceptual view of the hardware architecture [19]. We use the term *graphics processor*, or *accelerator*, to refer to a circuit comprising one or more logical compute devices. To avoid confusion between monolithic and multi-chip assemblies, we use AMD's term *graphics compute die* to refer to a group of compute units on a single semiconductor die that maps to a logical compute device [20]. We use the term *operational intensity* to refer to the ratio of computational work to communication in a program, for example, floating-point operations per byte accessed. We use the term *machine balance* to refer to the operational intensity required to reach both the computational and data transfer limits of the hardware, for example, the ratio of peak floating-point operations per second to the peak memory bandwidth.

## 2.2 | GPU architecture and execution model

Over the last two decades, GPUs have evolved from fixed-pipeline geometry processors to programmable accelerators [21,22] that excel in computationally intensive tasks, where the same instruction can be applied on multiple data elements in parallel [23]. Datacenter GPUs, while still graphics accelerators at heart [24], are designed to perform better under scientific workloads. They provide higher memory bandwidth and capacity, and double-precision throughput compared to consumer-grade GPUs tuned for real-time graphics [16,25,26,27].

A modern GPU is a multithreaded vector processor capable of executing a single instruction on multiple data items (SIMD). It consists of thousands of cores, where each core comprises a set of arithmetic-logic units and an allocation of the register file. A group of cores forms a compute unit (CU), where the cores have access to a shared L1 cache. To supply enough data to keep the massive number of parallel cores busy, the throughput of the GPU memory system has been maximized at the expense of off-chip memory access latency [24], which can be several hundred clock cycles on an L2 miss [28,29]. To hide memory access and operational latencies, a GPU programs must have sufficiently many instructions in flight. This has been alleviated with hardware multithreading, where groups of SIMD threads are multithreaded on the CUs [18,30]. The software abstraction of this group is called a thread block. In addition to thread-level parallelism (TLP) provided by hardware multithreading, instruction-level parallelism (ILP) has also been shown to be effective at hiding pipeline latencies [31].

Historically, Nvidia has been the leading supplier of HPC GPUs [32]. However, with the introduction of AMD GPUs on two of the three fastest supercomputers in 2022 [33,34], programs written for Nvidia's CUDA must be modified to make use of the new hardware. While the GPU architectures of both vendors are built upon similar ideas, there are some key differences.

The most notable difference between the devices of the two manufacturers is the cache configuration. On Nvidia Volta and newer GPUs, a portion of the L1 can be allocated as shared memory for cooperation between a group of threads operating on one CU [18,35]. On AMD CDNA 2 GPUs, shared memory is allocated on a separate memory unit, called local data share (LDS), which resides outside the CU [30]. Capacities of the memory regions also differ. For example, the shared memory capacity of an MI250X is roughly $2.5$ times smaller compared to the A100, but its computational FP64 performance per CU is roughly $2.4$ higher. Therefore, programs must achieve higher operational intensity with less shared memory capacity to reach machine balance on the MI250X. We refer the reader to Table 1 for a comparison of hardware specifications.

## 2.3 | Mathematical conventions

Throughout this work, we use the term *tensor* to refer to a multidimensional array [36]. This definition can be interpreted as a generalization of matrices to higher dimensions [37] and should not be confused with, for example, a stress tensor in physics, which carries additional meaning. The number of dimensions of a real-valued tensor $\mathcal{A}$ is denoted by its order $d$, i.e. $\mathcal{A} \in \mathbb{R}^{n_1 \times n_2 \times \cdots \times n_d}$. We denote the shape of a tensor as a $d$-tuple $(n_1, \ldots, n_d)$, where $n_i$ is the number of elements in dimension $i$. The index set w.r.t. dimension $i$ is written as $N_i = \{1, 2, \ldots, n_i\}$. The symbols $i$, $j$, $k$, and $w$ are used for index notation throughout and not assigned any specific meaning.

We refer to tensors of order $0$ as *scalars* denoted by lower-case italic symbol $a$. A tensor of order $1$ is denoted by a bold symbol $\mathbf{a} = [a_i]_{i \in N} = [a_1 \ \ldots \ a_n]$ and referred to as a *vector*. Finally, we use the term *matrix* to refer to a tensor of order $2$ denoted by uppercase bold symbol $\mathbf{A} = [a_{ij}]$, where $i \in N_1$ and $j \in N_2$. We use colons to denote subtensors. In the context of matrices, $\mathbf{a}_j$ is used to refer to the $j$th column vector and $\mathbf{a}_{i:}$ to the $i$th row vector. A submatrix of $\mathbf{A}$ composed of the rows $i$ to $j$ and columns $k$ to $w$ is denoted as $\mathbf{A}_{i:j,k:w}$. Finally, a tensor of order $3$ is denoted by bold uppercase italic symbol $\mathcal{A} = [a_{ijk}]$, where $i \in N_1, j \in N_2$, and $k \in N_3$. In general, the element $(i_1, \ldots, i_d)$ of a $d$-dimensional tensor is denoted by $a_{i_1 \ldots i_d}$, where $i_j \in N_j$. We use Iverson bracket notation [38]

$$[P] = \begin{cases} 1, & \text{if } P \text{ is true} \\ 0, & \text{otherwise.} \end{cases} \quad (1)$$

as a generalization of the Kronecker delta to simplify the construction of tensors. For example, the identity matrix $I = [\delta_{ij}]$ can be constructed by letting $\delta_{ij} = [i = j]$.

## 2.4 | Stencil computations

In numerical analysis, the term *stencil* is used to denote the visual representation of points, illustrated in Fig. 2, required to compute a numerical solution to a partial differential equation (PDE). However, the computational concept, where a fixed spatiotemporal neighborhood is used as input to a function that produces an output at the point of interest, is more general and can be found in a wide range of domains. For example, the nonzero values of a convolution kernel form a type of stencil. Other common applications include finite-difference solvers. In this work, we use the more general interpretation of stencils as the computational pattern instead of being strictly limited to PDEs.

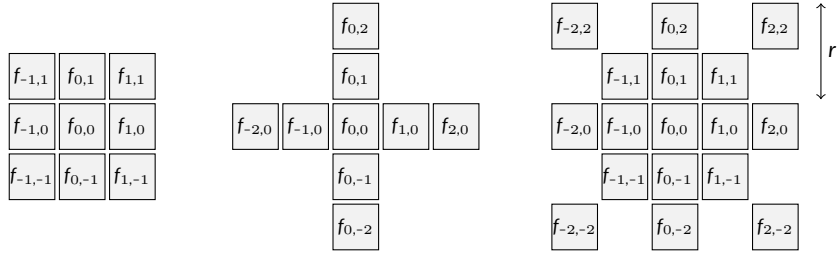

**FIGURE 2** Symmetric stencils centered around the point of interest at $(0, 0)$. The stencil influence radius is denoted as $r$. The underlying value at the index $(i, j)$, is denoted as $f_{i,j}$.

Stencil computations can be performed on both structured and unstructured grids. In a structured grid, the interior vertices form a regular subgraph, whereas in an unstructured grid, a vertex can have an arbitrary number of neighbors[39]. In direct numerical simulations based on finite differences, structured grids are a popular choice because they are space efficient, easy to implement with multidimensional arrays, and follow a straightforward indexing scheme[39]. Unstructured meshes are better suited for simulations in complicated geometric domains, such as modelling fluid flow in blood vessels, or when the flow develops very sharp structures, such as shocks in aircraft design[40]. In this work, we focus on computations on structured grids.

Stencil computations are typically performed in iterations, where an updated state $\mathbf{f}' \in \mathbb{R}^n$ is computed by applying a fixed stencil function $\zeta$ on each point of interest $i \in N$ using information constructed from the previous state as input. We present the one-dimensional case here for notational simplicity without loss of generality to multiple dimensions. We use the prime symbol to denote the updated state throughout this work. For derivatives, we use Leibniz's notation $\frac{d}{dx}$ to avoid confusion. We use the term *stencil influence radius*, or just *stencil radius*[6], to refer to the maximal Chebyshev distance from the point of interest to the stencil points required to evaluate $\zeta$. Formally, $r = \max_{s \in S} (\max_i |s_i - p_i|)$, where $S$ is an index set of the stencil points associated with the point of interest at index $(p_1, \ldots, p_d)$. This notion is useful, because all data required for the computations is guaranteed to fall within the influence radius, which simplifies the notation for the rest of this article while still applying to stencils with an arbitrary stride or displacement.

If the influence radius of the stencil is greater than zero, the stencil will extend beyond the computational domain when evaluating $\zeta$ near boundaries. To preserve the shapes of $\mathbf{f}$ and $\mathbf{f}'$ across iterations, $\mathbf{f}$ must be padded depending on the size of the stencil. For this, we introduce a boundary value function $\beta(\mathbf{f}, i)$, which can be used to construct an augmented array

$$\hat{f}_i = \begin{cases} f_i & \text{if } i \in N \\ \beta(\mathbf{f}, i) & \text{otherwise.} \end{cases} \quad (2)$$

The shape-preserving stencil iteration can now be written for arbitrary stencil functions as $f'_i = \zeta(\hat{\mathbf{f}}, i) \quad \forall i \in N$, where $\mathbf{f}, \mathbf{f}' \in R^n$. If the stencil is symmetric and centered around the point of interest $i$, $\hat{\mathbf{f}} \in \mathbb{R}^{n+2r}$.

When $\zeta$ produces an output that is a linear combination of its input, we refer to $\zeta$ as a linear stencil function, which is analogous to linear filtering in image processing. For example, consider discrete cross-correlation $\zeta(\hat{\mathbf{f}}, i) = \sum_{j=-r}^{r} g_j \hat{f}_{i+j}$ of real-valued functions $\mathbf{f}$ and $\mathbf{g}$, where $\mathbf{g}$ has finite support in $\{-r \ldots r\}$. Graphics processors offer hardware acceleration for the fused-multiply-add operation[41,20] reducing the number instructions required to compute functions of this form. Furthermore, by unrolling and reshaping the computations as a matrix-matrix product[42], hardware-accelerated matrix-fused-multiply-add units[16,43,20] can be leveraged to obtain further reductions in instruction counts. This is employed in the implicit general matrix multiplication algorithm, which is currently a standard technique for computing direct cross-correlations on graphics processors[43].

Nonlinear stencil functions are more challenging to compute because they cannot be solved with standard methods for matrix algebra. Examples of nonlinear stencil functions include max filtering $\zeta(\hat{\mathbf{f}}, i) = \max_{j \in \{-r \ldots r\}} \hat{f}_{i+j}$ and the numerical computation of nonlinear systems.

## 3 | METHODS

We investigate the performance of stencil computations in three cases. Firstly, we establish the baseline performance under ideal conditions in synthetic one-dimensional benchmarks, introduced in Section 3.1. Secondly, we benchmark linear stencil computations in one-, two-, and three-dimensions using the diffusion equation as the test case. The method for computing the diffusion equation as a linear stencil function is presented in Section 3.2. Finally, in the most challenging test, we study the performance of nonlinear stencil computations in three dimensions. We use the simulation of compressible magnetohydrodynamics as the test case, presented in Section 3.3.

## 3.1 | Cross-correlation

We use one-dimensional discrete cross-correlation over real vector spaces as the stencil function in our baseline performance tests. The updated state is computed as

$$f'_i = \sum_{j=-r}^{r} g_j \hat{f}_{i+j} \,. \tag{3}$$

This represents a nearly ideal computation task for graphics processors, because each point of interest $i \in N$ can be updated in parallel using a fixed function, the memory locations required to update nearby elements are highly localized, enabling high cache hit rate, and hardware-accelerated fused-multiply-add can be used to accumulate the result. Furthermore, one-dimensional linear stencil computations on a single array avoid the tuning challenges of more complex, multi-dimensional stencil computations.

## 3.2 | Diffusion equation

In the second set of tests, we establish the performance of linear stencil algorithms by computing the numerical solution to the diffusion equation in one to three dimensions. The computation of partial differential equations is particularly interesting, because it lacks a native mapping to the graphics processor hardware in terms of graphics or machine learning primitives, but can still benefit greatly from improved memory throughput and parallelism.

The diffusion equation $\frac{\partial f}{\partial t} = \alpha \nabla^2 f$ with a constant diffusion coefficient $\alpha$ is a linear system, which can be solved numerically, for example, with the forward Euler method for time integration as $f'_i = f_i + \Delta_t \alpha \nabla^2 f_i$. Here $f'_i$ denotes the updated state, $\Delta_t$ the time step length, and $\nabla^2 f = \sum_{i=1}^{d} \frac{\partial^2 f}{\partial x_i^2}$ the Laplacian for $f$ in $d$ dimensions. By substituting $f_i = \sum_{j=-r}^{r} c_j^{(1)} \hat{f}_{i+j}$, where $c_j^{(1)} = [j = 0]$ and $\frac{\partial^2 f_i}{\partial x^2} = \sum_{j=-r}^{r} c_j^{(2)} \hat{f}_{i+j}$, where $\mathbf{c}^{(2)}$ is the vector of central difference coefficients for the second derivative, the system can be solved numerically in one dimension by

$$f'_i = \sum_{j=-r}^{r} c_j^{(1)} \hat{f}_{i+j} + \Delta_t \alpha \sum_{j=-r}^{r} c_j^{(2)} \hat{f}_{i+j} \tag{4}$$

$$= \sum_{j=-r}^{r} \left( c_j^{(1)} + \Delta_t \alpha c_j^{(2)} \right) \hat{f}_{i+j} \,. \tag{5}$$

This can be expressed as a discrete cross-correlation $\mathbf{f}' = (\mathbf{g} \star \hat{\mathbf{f}})$ where $\mathbf{g} = \left[ c_j^{(1)} + \Delta_t \alpha c_j^{(2)} \right]_{j \in \{-r \ldots r\}}$. Because cross-correlation is distributive over addition for real-valued functions, adjusting the computations to multiple dimensions is straightforward by letting $\nabla^2 f_i = \frac{\partial^2 f_i}{\partial x^2} + \frac{\partial^2 f_i}{\partial y^2} + \ldots$ and writing the state update in the form

$$f'_i = \left( \mathbf{g}^{(1)} \star \hat{\mathbf{f}} \right)_i + \left( \mathbf{g}^{(2)} \star \hat{\mathbf{f}} \right)_i + \left( \mathbf{g}^{(2)} \star \hat{\mathbf{f}} \right)_i + \ldots \tag{6}$$

$$= \left( \left[ \mathbf{g}^{(1)} + \mathbf{g}^{(2)} + \mathbf{g}^{(3)} + \ldots \right] \star \hat{\mathbf{f}} \right)_i \,. \tag{7}$$

## 3.3 | Magnetohydrodynamics

As the final set of tests, we explore the performance characteristics of nonlinear stencil computations in simulations of turbulent, compressible plasma. In contrast to our previous test cases, there are now several interacting fields. Throughout this work, we use the term *field* to refer to a physical field or quantity, such as a scalar density field, or to one component of a vector field, such as of a 3D velocity field.

The underlying equations in plasma simulations are based on combining the Navier-Stokes equations of fluid motion with the equations of electromagnetism. The study of such systems is called magnetohydrodynamics. This nonlinear system of equations has no known analytical solution in the non-trivial regime of developed turbulence, but has to be solved numerically. It is a realistic example of a typical scientific problem fit to be solved on a graphics processors.

Our setup is based on non-ideal magnetohydrodynamic equations in the non-conservative form, presented in Appendix A. The simulation domain consists of two scalar fields, density and entropy, and two vector fields, velocity and magnetic vector potential. The fields are coupled with each other, with the ideal gas law utilized to provide a closure between the thermodynamic quantities. The state of the simulation is advanced in time using explicit Runge-Kutta three-time integration and spatial derivatives are computed using 6th-order central differences. Therefore, an integration step comprises three passes over the elements, where spatial derivatives are computed using radius three stencils. For more details on the physical aspects of the setup and motivation for the selection of the integration and differentiation methods, we refer the reader to [44].

We can represent the simulation loop as a composition of functions $\phi(\gamma(\psi(\mathbf{f})))$, where $\phi$ is a nonlinear transformation, $\gamma$ a linear transformation, and $\psi$ a function that pads the spatial dimensions of the input $\mathbf{f}$. This formulation is inspired by *feedforward neural networks*, where nonlinearity is achieved by applying a nonlinear transformation $\phi$ on neurons mapped with a linear transformation[45]. In fact, the sequence of tensor transformations used to advance the simulation bears remarkable similarity to the structure of convolutional neural networks, where each transformation can be viewed as a layer in an infinitely deep feedforward network.

We can represent the discretized simulation domain as a matrix $\mathbf{F}$ that is constructed by concatenating $n_f$ scalar arrays corresponding to the components of the physical fields written as column vectors. Likewise, the coefficients of $n_s$ cross-correlation kernels can be arranged in a matrix $\mathbf{A} = [a_{ij}] \in \mathbb{R}^{n_s \times n_k}$, where $\mathbf{a}_{i:}$ is a row vector corresponding to the $i$th cross-correlation kernel and $n_k$ the size of the kernel. To compute the output with a symmetric stencil centered around the point of interest $i$, as is the case with central differences, it suffices to consider the submatrix $\mathbf{B}^{(i)} = \hat{\mathbf{F}}_{i-r:i+r,:} \in \mathbb{R}^{n_k \times n_f}$, where $n_k = 2r + 1$ and $r$ the stencil radius.

With these formulations, we can now write

$$\gamma\left(\mathbf{B}^{(i)}\right) = \mathbf{A}\mathbf{B}^{(i)}, \tag{8}$$

where $\gamma : \mathbb{R}^{n_k \times n_f} \to \mathbb{R}^{n_s \times n_f}$. The nonlinear output can then be computed with $\phi : \mathbb{R}^{n_s \times n_f} \to \mathbb{R}$. If the input tensor has more than one spatial dimension, the values in the spatial dimensions of the subtensor $\mathcal{B}$ can be flattened to fit a single column, adjusting the coefficient matrix $\mathbf{A}$ accordingly. For updating multiple scalar arrays, the state update becomes

$$f'_{ij} = \phi_j\left(\gamma\left(\mathbf{B}^{(i)}\right)\right), \tag{9}$$

where $\phi_j$ is the function to update the $j$th array.

# 4 | IMPLEMENTATIONS

## 4.1 | Cross-correlation using handcrafted CUDA and HIP programs

We created a small program using CUDA and HIP to measure baseline performance. The purpose of this program is to serve as a clean-room experiment that can achieve near-ideal hardware utilization. Two caching strategies were implemented: one that relies on *hardware-managed caching (HWC)* for data reuse, and one where data reuse is primarily achieved by *software-managed caching (SWC)*. In HWC, the contents of the L1 and L2 caches are influenced by the cache replacement policy implemented in hardware, whereas in SWC, the contents of shared memory are controlled by the programmer in addition regular HWC[46].

In our HWC implementation, we assign the threads of a thread block to subsequent outputs in the one-dimensional output tensor. Because the threads process the stencil elements in a fixed order from left to right, the memory accesses of a warp are guaranteed to fall on a contiguous block of memory. This approach ensures spatial locality of the memory accesses, and results in improved memory access coalescing, because few redundant memory transactions are required across 32-byte data blocks[28].

The drawback of HWC is that there are limited ways to influence the data placement in the cache hierarchy due to thousands of concurrent threads competing for the same resources. Hence, explicit cache management is often recommended as a tuning strategy for improving on-chip reuse[18]. Our approach to SWC is otherwise the same as with HWC, except the full working set is fetched into shared memory before iterating over the stencil elements. Because each thread accesses a separate bank at each iteration, this approach is free of bank conflicts.

These implementations alone were not enough to attain bandwidth-bound performance is some of the tests, primarily when using single precision. To remedy this, we applied two additional tuning strategies for reducing latencies: computing multiple outputs per thread in the innermost loop, and unrolling the computations of the multiply-accumulate loop over the stencil points. We refer to these approaches as *element-wise* and *stencil point-wise unrolling*, respectively. For unrolling both approaches, we used the `#pragma unroll` intrinsic. C++ templates were used to ensure the loop increment and bounds were known at compile time. To ensure a fair comparison between the systems, we used automated tuning for extracting the optimal domain decomposition and refrained from applying architecture- or vendor-specific optimizations.

## 4.2 | Cross-correlation using cuDNN and MIOpen

We implemented the baseline tests also with cuDNN[47] and MIOpen[48] to ground the results in state-of-the-art implementations. We used the two-dimensional variant of the convolution function and ordered the data in the channels-last format (NCHW). NCHW refers to a data layout of a four-dimensional tensor with two spatial dimensions, where, from slowest to the fastest moving dimension, N is the batch size, C the number of channels, H the height, and W the width. We chose NCHW, because it gave better performance in one-dimensional convolutions, where the batch size and number of channels was one, compared to the channels-first format (NHWC), which is typically recommended for machine learning applications[43]. With both libraries, we selected the most efficient algorithm with an exhaustive search using functions `cudnnFindConvolutionForwardAlgorithm` and `miopenFindConvolutionForwardAlgorithm`.

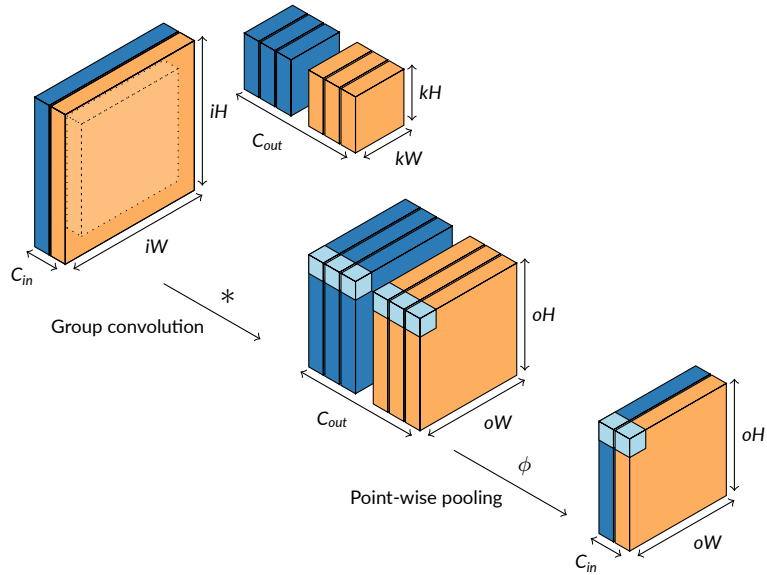

**FIGURE 3** An illustration of tensor transformations used to compute the diffusion and MHD equations with PyTorch. Here the input is a two-channel tensor $(1, C_{in}, oH, oW)$ in NCHW format where the spatial dimensions have been padded to $(iH, iW)$. A group convolution is applied to both channels using three filters, resulting in an output tensor with six channels and shape $(1, C_{out}, oW, oH)$. Finally, a nonlinear operation $\phi$ is applied to each spatial point in the tensor, reducing $C_{out}$ channels back to $C_{in}$ channels.

## 4.3 | Diffusion equation and magnetohydrodynamics using PyTorch

As a proof of concept, we implemented the chain of tensor transformations to compute the diffusion and MHD equations using PyTorch. PyTorch uses cuDNN and MIOpen as a backend, but provides a convenient high-level interface for implementing more challenging computations. For computing one- to three-dimensional cross-correlations in the diffusion and MHD equations, we used the family of `torch.nn.functional.conv` functions provided by PyTorch. Conceptually, we implemented the computations as $n_s$ convolution filters applied separately on each channel of the input tensor, where each channel corresponds to one of the $n_f$ arrays that depict the components of the physical fields. Nonlinearity is achieved by applying point-wise pooling over channels based on $\phi$. This approach is visualized in Fig. 3.

In practice, we implemented the forward pass by distributing fields to different batches due to PyTorch being the only library that supports grouped convolutions at the time of writing. We did not notice a performance difference between group and batched convolutions in our initial tests and did not pursue further comparisons. We chose NCHW as the data layout based on our tests with cuDNN and MIOpen. When implementing and optimizing the computations, we followed the tuning and benchmarking guidelines of PyTorch. Notably, we used just-in-time compilation to fuse and compile the forward pass, and enabled automated tuning with the `torch.backends.cudnn.benchmark` module. Furthermore, we ran the benchmarks in inference mode, which disables automatic differentiation and other operations required only for model training. Tensor cores were disabled, because mixed-precision computations resulted in a loss of accuracy outside our error tolerance. For benchmarking, we used the `torch.utils.benchmark` module to ensure proper timing and synchronization.

## 4.4 | Diffusion equation and magnetohydrodynamics using Astaroth

We used the Astaroth framework[49,50,51] for evaluating hardware- and software-managed caching in more challenging computations. Astaroth is tuned for nonlinear stencil computations with a special focus on enabling large-scale simulations of magnetohydrodynamics. The framework consists of a software library, a domain-specific language (DSL), and a code generator. The library implements communication and scheduling for distributed computations[6,52], while the code generator translates high-level problem descriptions written in the DSL into optimized CUDA kernels embedded in the library[49]. For this work, we have redesigned both the code generator and the DSL to enable the optimization techniques presented here, and added support for AMD devices.

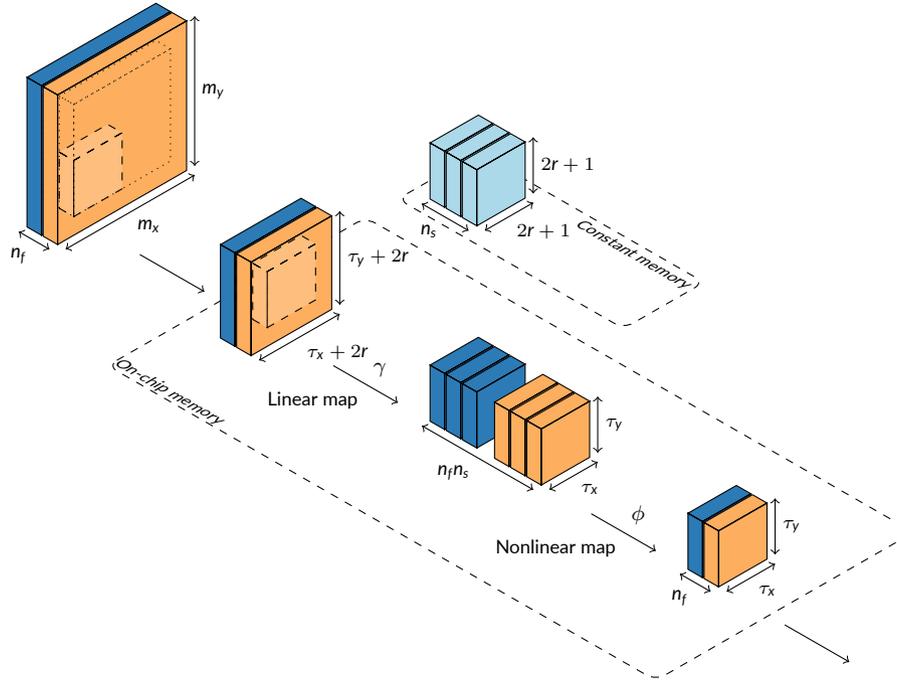

**FIGURE 4** An illustration of tensor transformations used to compute the diffusion and MHD equations with Astaroth. The concept is largely the same as in Fig. 3, but presented here with a more detailed view of the internal computations. With Astaroth, the input is processed in blocks of $(\tau_x, \tau_y, \tau_z)$ grid points, where each thread produces one spatial point as output.

Our code generator implements operator fusion for computations that can be expressed in the form $\phi(\mathbf{AB})$. Operator fusion is an optimization technique, where redundant accesses to off-chip memory are eliminated by performing a chain of operations within a single kernel invocation. The generated kernel follows a structure, where the result of $\mathbf{AB}$ is computed on-chip and placed in local memory, where it is reused to compute multiple nonlinear outputs. This strategy, illustrated in Fig. 4, is especially effective in multiphysics applications, where information is required from all of the interacting fields to compute the state update.

The set of linear stencil functions used to compute $\phi$ can be defined with language constructs provided with the DSL. At compile time, this information is used to deduce the shapes of $\mathbf{A}$ and $\mathbf{B}$. At runtime, $\mathbf{A}$ is placed in device constant memory and a thread-local subtensor $\mathcal{B}$ is constructed from the input tensor $\mathcal{F}$ residing in off-chip memory. The input tensor $\mathcal{F}$ is stored in a row-wise scan pattern in memory, where the spatial index $(i, j, k)$ maps to a linear index $i + jn_x + kn_xn_y$, where $(n_x, n_y, n_z)$ are the spatial dimensions of $\mathcal{F}$. In three-dimensional computations, we flatten the spatial dimensions of the thread-local subtensor $\mathcal{B}$ in a scan pattern over $x$, $y$, and finally, $z$ spatial axes to obtain $\mathbf{B} = [\mathbf{b}_j]_{j \in N_f}$, where the column vector $\mathbf{b}_j$ corresponds to the flattened input elements associated with the $j$th array. The construction of $\mathbf{A} = [\mathbf{a}_{i:}]_{i \in N_s}$ follows the same pattern, where the row vector $\mathbf{a}_{i:}$ corresponds to the flattened coefficients of the $i$th kernel.

The performance characteristics of the generated code can be tuned by changing the order of computation of $\mathbf{AB}$, changing the number of outputs each thread computes, and choosing memory placement for $\mathbf{A}$ and $\mathbf{B}$. In this work, we evaluate two approaches for storing the data. In both approaches $\mathbf{A}$ is stored in constant memory, whereas $\mathbf{B}$ is stored either explicitly in shared memory or implicitly in the hardware-managed cache hierarchy. Furthermore, each GPU thread is assigned a point of interest, where it computes multiple outputs corresponding to the updated states of the physical fields.

In our first implementation, we let the hardware choose the placement of $\mathbf{B}$ in the memory hierarchy, and evaluate the matrix product $\mathbf{AB}$ in a tiled fashion by iterating over the columns of $\mathbf{B}$ first. The visual representation of this approach is showcased in Fig. 5a. With this evaluation order, subsequent instructions are guaranteed to be independent, reducing stalls due to data dependencies. Deeper instruction pipelining can be achieved by increasing the size of the column tile, however, this also increases the size of the working set and can lead to cache thrashing. The general order for computing $\mathbf{AB}$ is shown in Algorithm 1. In our implementation, we unroll these loops completely during code generation to reduce overhead from integer arithmetic and control flow. Furthermore, we prune instructions that do not contribute to the result, i.e. operations where

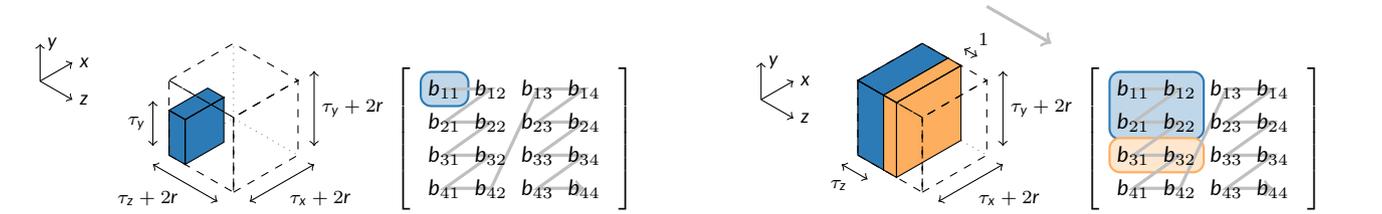

(a) An illustration of the HWC implementation. The memory locations accessed during the initial computation step are colored blue. The tensor view depicts the spatial memory accesses of a thread block, whereas the matrix view depicts the memory accessed by a single thread.

(b) An illustration of the SWC implementation. The memory locations available in shared memory are colored blue. The prefetch buffer updated in parallel with computations is colored orange.

**FIGURE 5** Tensor and matrix views of the hardware- and software-cached implementations. The gray arrow shows the general direction of computations. The three-dimensional tensor views illustrate the spatial arrangement of the data and how it is visible to the thread block. Dashed lines illustrate the extents of the working set. The matrix views illustrate how the spatial data is accessed by a single thread. For simplicity and space considerations, the tensor view depicts the spatial extents of only a single column in the matrix view, whereas in the matrix view, the spatial dimensions have been flattened to span the rows, and two columns are being operated on.

the stencil coefficient is zero or associated with an unused cross-correlation. This feature is enabled by compiling the Astaroth library with the `OPTIMIZE_MEM_ACCESSES` option.

---

**Algorithm 1** The evaluation order for computing **AB**. For simplicity, blocking w.r.t. $N_f$ has been left out.

Data: Coefficient matrix $\mathbf{A} \in \mathbb{R}^{n_s \times n_k}$ and input matrix $\mathbf{B} \in \mathbb{R}^{n_k \times n_f}$.
Result: Linear map $\mathbf{Q} = \mathbf{AB}$.
  $\mathbf{Q} \leftarrow [0]_{n_s \times n_f}$
  for $k = 1$ to $n_k$ do
    for $i = 1$ to $n_s$ do
      for $j = 1$ to $n_f$ do
        $q_{ij} \leftarrow q_{ij} + a_{ik} b_{kj}$
      end for
    end for
  end for

---

In our other implementation, showcased in Fig. 5b, we store **B** into shared memory. However, the primary challenge in three-dimensional multiphysics applications with large stencils is that the subtensor required by a thread block, which is sufficiently large for achieving an adequate on-chip reuse ratio, does not fit into shared memory on current hardware [‡]. In our approach, we work around this by allocating a block of $(\tau_x + 2r, \tau_y + 2r, \tau_z)$ elements in shared memory, which we stream along the z-axis until all of the $(\tau_x + 2r, \tau_y + 2r, \tau_z + 2r)$ spatial input elements have been processed. To enable pipelining of memory fetches and compute, we also allocate a prefetch buffer of $(\tau_x + 2r, \tau + 2r, 1)$ elements and use it to update the leading edge of the streamed data block. We implement the shared memory allocation as a circular buffer to eliminate the need for shuffling the data when updating the leading edge. To reduce stalls due to data dependencies among subsequent instructions, we process multiple columns successively in the same fashion as with the hardware-cached approach. In the benchmarks of this work, we processed four components of the physical fields in shared memory at a time. Lastly, to reduce the arithmetic overhead arising from index calculations required for fetching the data into shared memory, we precompute and unroll the fetches by a factor of four under the requirement there are enough threads such that the shared memory block can be updated with $\leq 4$ fetches per thread. The benefit of this approach is that it supports both two- and three-dimensional cache blocking and the optimal decomposition can be searched at runtime. This is useful, because the ideal decomposition is difficult to choose based on hardware specifications alone due to the complex interaction between various factors affected by the decomposition, such as register pressure, occupancy, and data reuse factor.

We added support for AMD devices using the heterogeneous interface for portability (HIP)[17], which is a compatibility layer that provides near one-to-one mapping with CUDA functions. For this, we created a header file containing preprocessor macros for translating all CUDA calls to their HIP equivalents, and compiled the library with the HIP compiler.

---

[‡] For example, a block of $(\tau_x, \tau_y, \tau_z) = (8, 8, 8)$ threads requires an input subtensor $\mathcal{B}$ that contains $n_f(\tau_x + 2r)(\tau_y + 2r)(\tau_z + 2r) = 21\,952$ elements with $r = 3$ and $n_f = 8$. This is $\approx 172$ KiB when using double precision, whereas AMD Instinct MI200 and MI300 series accelerators provide only $64$ KiB of shared memory[53].

**TABLE 1** A list of graphics processing units used in this work. On AMD graphics processors, L1 and shared memory transactions are serviced from separate memory spaces, whereas on Nvidia, the L1 cache and shared memory are unified and the allocation of software-managed memory region reserved from L1 can be chosen at runtime. In this table, we list the maximum allocation of shared memory that can be reserved from the unified L1 cache.

| Description | A100 [16,18] | V100 [54,29,16,18] | MI250X [20,30] | MI100 [26,55,56] |
|---|---|---|---|---|
| Model | SXM4-40GB | SXM2-32GB | HBM2e OAM | HBM2 PCIe |
| Vendor | Nvidia | Nvidia | AMD | AMD |
| Release year | 2020 | 2018 | 2021 | 2020 |
| SIMD width | 32 | 32 | 64 | 64 |
| Graphics compute dies (GCDs) | 1 | 1 | 2 | 1 |
| Compute units (CUs) per GCD | 108 | 80 | 110 | 120 |
| FP32 cores per GCD | 6912 | 5120 | 7040 | 7680 |
| FP64 cores pre GCD | 3456 | 2560 | 7040 | - |
| Compute clock (MHz) | 1410 | 1530 | 1700 | 1502 |
| Peak vector performance per GCD (FP64 TFLOPS) | 9.7 | 7.8 | 23.9 | 11.5 |
| Machine balance (FP64 FLOPS/8 bytes) | 50 | 70 | 117 | 75 |
| L1 cache per CU (KiB) | 192 | 128 | 16 | 16 |
| L2 cache per GCD (MiB) | 40 | 6 | 8 | 8 |
| Shared memory per CU (KiB) | 164† | 96† | 64 | 64 |
| Memory clock (MHz) | 1215 | 877 | 1600 | 1200 |
| Memory capacity per GCD (GiB) | 40 | 32 | 64 | 32 |
| Memory bandwidth per GCD (GiB/s) | 1448 | 835 | 1526 | 1144 |
| Thermal design power | 400W | 300W | 560W | 300W |

†Carved from L1.

**TABLE 2** A list of systems and software used in this work. The programs used for this work are available in the Astaroth repository [51]. The diffusion and MHD equation benchmarks with Astaroth were generated with commit `b6c36c4`. Benchmarks showcasing the performance with CUDA, HIP, cuDNN, and MIOpen were generated with commit `51abfaa`. PyTorch benchmarks were generated with commit `99b6a20`.

| Specification | Mahti | Puhti | LUMI | Triton |
|---|---|---|---|---|
| CPU | 2× AMD Rome 7H12 | 2× Xeon Gold 6230 | AMD EPYC 7A53 | 2× AMD EPYC 7262 |
| GPU | 4× Nvidia A100 | 4× Nvidia V100 | 4× AMD MI250X | 3× AMD MI100 |
| CUDA/ROCm | 11.5.0/- | 11.2.2/- | -/5.2.3 | -/5.0.0 |
| cuDNN/MIOpen | 8.3.3.40/- | 8.0.5.39/- | -/2.17.0 | -/2.15.0 |
| PyTorch | 2.2.1+cu121 | 2.2.1+cu121 | 2.2.1+rocm5.6 | 1.1 |

# 5 | RESULTS

## 5.1 | Setup

We benchmarked the graphics processors listed in Table 1 on systems listed in Table 2. It should be noted that the GCDs of the MI250X map to separate logical graphics processing units with their own memory space. Therefore, programs must be crafted with multi-device communication in mind to utilize the full accelerator. To avoid communication overhead and to enable the use of unified programs for benchmarking, we utilize only a single GCD of the two-die AMD MI250X throughout this work.

In the benchmarks, we randomized the input tensor, called the kernel several times as a warm-up, determined optimal thread block dimensions using automated tuning if applicable, and finally, measured the median running time of $100$ iterations. The time to pad was not benchmarked. We used a problem size of $64$ MiB for all single-precision tests and $128$ MiB for all double-precision tests, except the initial test, which we used to determine the effective memory bandwidth.

We verified the results by comparing the output tensors with a model solution. In CUDA, HIP, cuDNN, and MIOpen benchmarks, we asserted that the comparison is exact. We verified the Python implementations using function `allclose` provided with the `numpy` package, which we used to confirm that $|a - b| \leq (c + c|b|)$ for all elements in the output, where $a$ and $b$ are the ground truth and measured values, respectively, and $c$ the error tolerance. In the diffusion and MHD equation tests we used $c = 5\epsilon$ and $c = 100\epsilon$, respectively, where $\epsilon$ is the machine epsilon. With Astaroth, we asserted that the relative error is $< 5\epsilon$ or the absolute error less than the minimum value in the domain scaled to $\epsilon$. We refer the reader to Table B2 in Appendix B for more details on the setup. The runs for computing the MHD benchmarks were not verified due to the prohibitively long time to compute the model solution. Instead, we verified the implementations in smaller, $32^3$-point runs decoupled from the benchmarks.

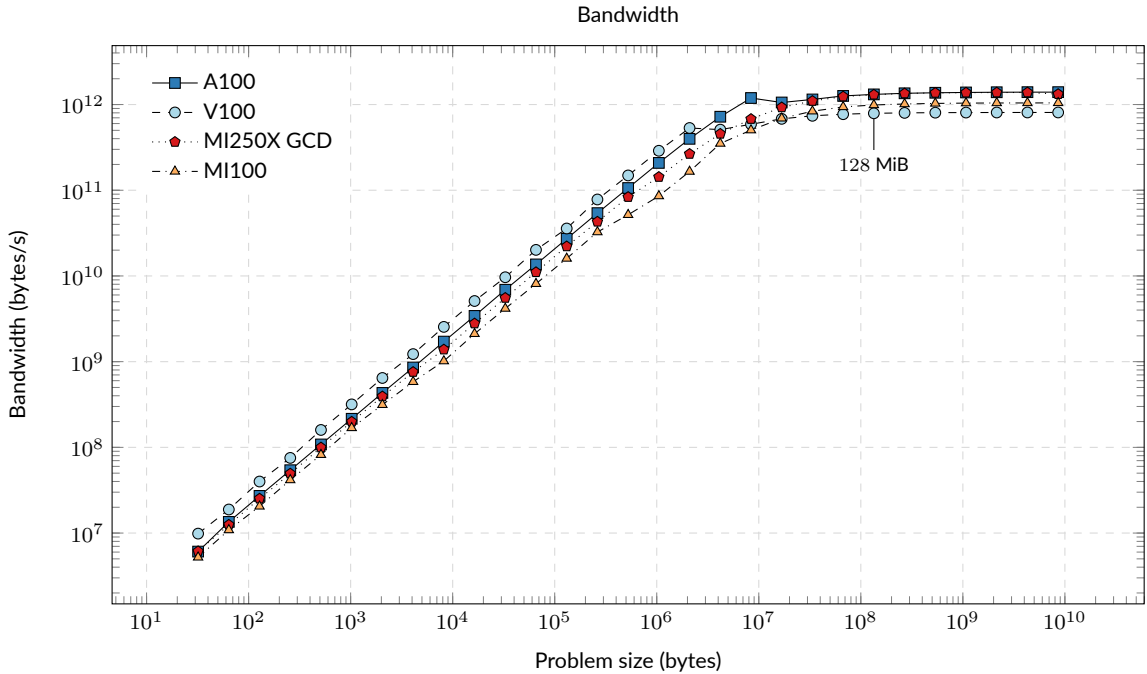

**FIGURE 6** Effective off-chip memory bandwidth measured with the CUDA/HIP implementation using double precision. We used the benchmark to determine the minimum problem size at $r = 0$ that is required to saturate the graphics processor with work and reach the ceiling of the effective memory bandwidth.

Automated tuning was performed on the CUDA/HIP kernels via a heuristic search of the valid combinations for the thread block dimensions ($\tau_x, \tau_y, \tau_z$). We pruned the search space by assuming that $\tau_x$ is a multiple of L2 cache line size divided by the size of double, which is $64/8 = 8$ on current Nvidia devices [29], and the optimal thread count per block was a multiple of the device's warp size. This is justified, because threads accessing subsequent addresses in the $x$ axis fall unto a single, or at most two, cache lines due to indexing the tensor in a row-wise scan (global memory transactions are 32 bytes on compute-capability 6.0 devices [28]). Furthermore, when the thread count is a multiple of warp size, the thread scheduler can dispatch instructions to all available streaming processors per cycle without having to mask a part of the threads inactive [18]. Decompositions that resulted in a failed launch, for example due to overusing on-chip resources, were discarded. Decompositions that resulted in a valid launch were timed over 3 iterations and the best-performing decomposition was selected for further benchmarking.

## 5.2 | Cross-correlation

In the first test, we measured the effective memory bandwidth using the CUDA/HIP implementation for cross-correlations with $r = 0$ equivalent to computing $f'_i = \zeta(\hat{\mathbf{f}}, i) = \hat{f}_i$. The results, shown in Fig. 6, were used to find the problem size large enough to hide timing overhead and saturate the memory bus. The effective bandwidths were $90\%$ (A100), $90\%$ (V100), $84\%$ (MI250X), and $85\%$ (MI100) of the theoretical maximum with double precision and $87\%$ (A100), $88\%$ (V100), $78\%$ (MI250X), and $79\%$ (MI100) with single. The minimum problem size for achieving $\geq 85\%$ utilization of the effective memory bandwidth on all devices with both single and double precision was $64$ MiB. We chose $64$ MiB as the problem size for further benchmarks with single precision § and $128$ MiB with double ¶.

In the next test, shown in Fig. 7, we measured the performance of one-dimensional cross-correlations computed with cuDNN [57] and MIOpen [48]. The speedups with an A100 over an MI250X GCD fell in range $2.3$ – $3.2$ with a median of $2.8$. With hand-tuned CUDA/HIP implementations, shown in Fig. 8, the performance gap between AMD and Nvidia devices was smaller. The speedups with an A100 over MI250X fell in range $1.0$–$1.8$ with a median of $1.5$ when using hardware caching and double precision. With SWC, the speedup with an MI250X GCD over A100 was in range $1.0$–$1.4$ with a median of $1.1$ with both single and double precision. At $r = 1024$, the best hardware-cached implementations were at most factors $1.03$ (A100), $1.13$ (V100), $1.88$ (MI250X), and $1.72$ (MI100) slower than the software-cached implementation. Despite being clearly slower in benchmarks with cuDNN/MIOpen, the MI250X GCD outperformed or was on par with other devices when using software-managed memory. The best CUDA implementation was $1.6 - 3.9$ times faster than cuDNN convolution on Nvidia devices. On AMD devices, the best HIP implementation was a factor $5.3 - 10.6$ faster than the MIOpen implementation.

---

§ Effective bandwidth utilization at 64 MiB with single precision: $90\%$ (A100), $96\%$ (V100), $88\%$ (MI250X), $91\%$ (MI100)
¶ Effective bandwidth utilization at 128 MiB with double precision: $94\%$ (A100), $98\%$ (V100), $94\%$ (MI250X), $95\%$ (MI100)

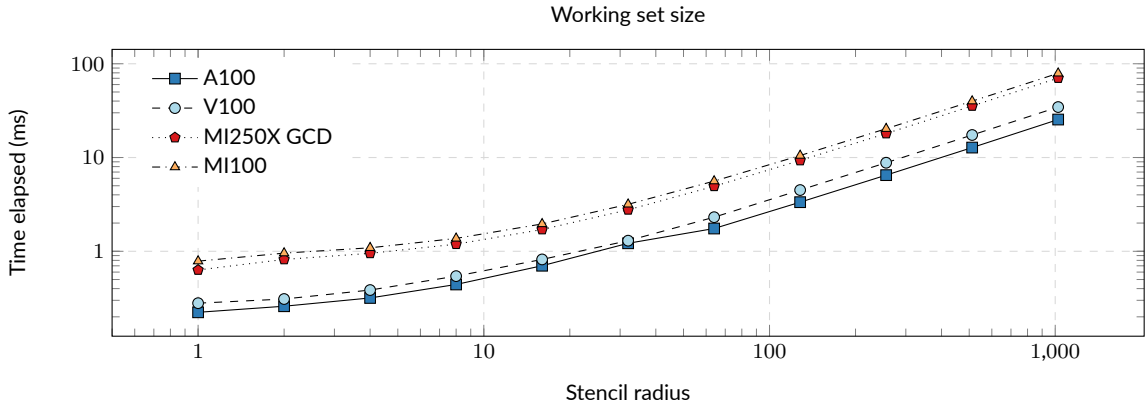

**FIGURE 7** Time per step of one-dimensional cross-correlation implemented with cuDNN/MIOpen using single precision. Lower is better.

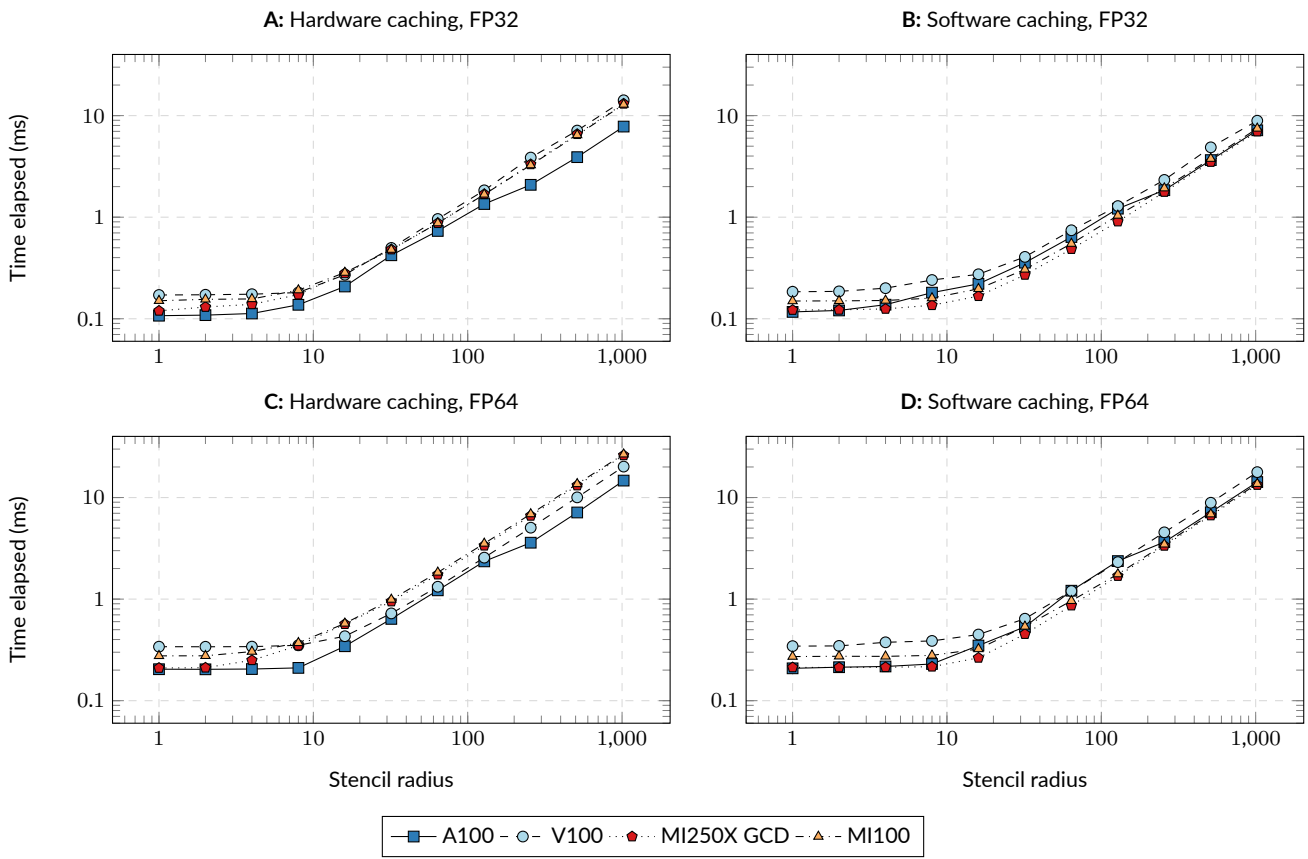

**FIGURE 8** Time per step of one-dimensional cross-correlation implemented with CUDA/HIP using the best-performing implementation for each device. Lower is better. The number of transactions served from L1/shared memory increase with stencil radius. Off-chip memory bandwidth limits the performance when the stencil radius is relatively small. At larger stencil radii, we saw indications that L1/shared memory bandwidth becomes saturated and starts to limit performance.

Fig. 9 shows a comparison of tuning strategies applied to reach the hardware limits. On an A100 with HWC and $r \geq 10$, profiling revealed $\geq 95\%$ throughput of the L1/TEX cache, indicating that the performance-bound was cache bandwidth. Stencil point-wise unrolling caused a clear performance pitfall on the MI100 and MI250X using FP32 in Fig 9F, whereas the effect subsided using FP64 in Fig 9L. Furthermore, element-wise unrolling shown in Fig. 9B and Fig. 9H was not an effective strategy on the MI100 and MI250X. Compared to baseline performance with hardware caching, speedups 3.1 (A100), 3.1 (V100), 2.7 (MI250X), and 2.7 (MI100) could be achieved with further tuning using FP32, and 1.6 (A100), 1.8 (V100), 3.9 (MI250X), and 3.9 (MI100) using FP64.

## 5.3 | Diffusion equation

In the next set of test, we benchmarked the numerical computation of the diffusion equation using PyTorch and Astaroth. The performance obtained with PyTorch is shown in Fig. 10. The performance of PyTorch compared to cuDNN/MIOpen is shown in Table C3 in Appendix C. The performance of 3D convolution at $r = 2$ on the MI250X degraded dramatically and the data point at $r = 2$, corresponding to $1800$ ms, is left out for clarity. We also obtained indications of the performance pitfall on the MI100 in a similar test with an earlier version of our benchmark program. In further tests on the MI250X with smaller problem sizes comprising only $128^3$ data points, the performance pitfall subsided, and the running time scaled to $r = 4$ as expected.

Fig. 11 shows the time per step for computing the diffusion equation with Astaroth. With single precision, the differences between the devices were minor, whereas with double precision, the A100 and V100 scale more efficiently to larger stencil radii. A similar effect is visible in the in the bottom left image of Fig. 8, where the A100 and V100 can sustain larger working sets without performance degradation than the MI100 and MI250X. A comparison of the HWC and SWC tuning strategies with Astaroth are shown in Fig. 12. It should be noted that the software-cached implementation was designed for MHD simulations, and does not leverage optimization techniques designed specifically for solving diffusion equation-like problems.

In addition to automated tuning at runtime, the implementations were tuned at compile time by changing the register allocation per kernel using the `__launch_bounds__` qualifier. In all cases, the default configuration without `__launch_bounds__` resulted in optimal register allocation. The results are shown in Fig. C1 in the Appendix C.

## 5.4 | Magnetohydrodynamics

In the most challenging set of tests, we benchmarked the time to compute the MHD equations in three dimensions using sixth-order finite differences, i.e. radius $3$ stencils, using PyTorch and Astaroth. The times to compute an integration substep with PyTorch were 41.9 (A100), 53.4 (V100), and 97.0 (MI250X) milliseconds.

Fig. 13 illustrates a comparison of tuning strategies implemented with Astaroth. The hardware-cached implementation was $1.8–2.9$ times faster than the software-cached implementation with single precision and $2.4–8.1$ times faster with double. Further inspection revealed that while the average stall incurred from long scoreboard dependencies [#] was reduced from $3.6$ cycles to $3.0$ cycles with the use of shared memory, the overall instruction count was increased $2.3$-fold due to additional index calculations required for managing the cache. Both implementations executed roughly the same amount of instructions per clock, $0.94$ with the HW-cached and $0.91$ with the SW-cached approach. During development, we encountered an unexpected performance pitfall resulting in a factor $6$ slowdown on AMD graphics processors when writing the result back to off-chip memory within a conditional expression depending on the value of a device constant. As a workaround, we wrote the conditional as an equivalent arithmetic expression that evaluated to zero in the `else` case. This workaround is enabled in all benchmarks.

Fig. 14 illustrates an exploration of the tuning parameters. In contrast to the diffusion equation benchmarks, the register allocation had to be manually tuned to achieve the highest performance on the MI100 and MI250X. Our MHD implementation achieved $19.6\%$ (A100), $17.9\%$ (V100), $10.5\%$ (MI250X), and $10.1\%$ (MI100) of ideal performance, where the computational domain would be read and written exactly once utilizing the peak theoretical bandwidth.

Finally, the performance per watt measurements of key benchmarks have been collected in Table 3. The MI250X GCD provided the best performance per watt for one-dimensional cross-correlations, whereas the A100 was the most energy-efficient in three-dimensional MHD simulations.

---

[#] On Nvidia hardware, long scoreboard stalls arise from local, global, surface, or texture operations [58]

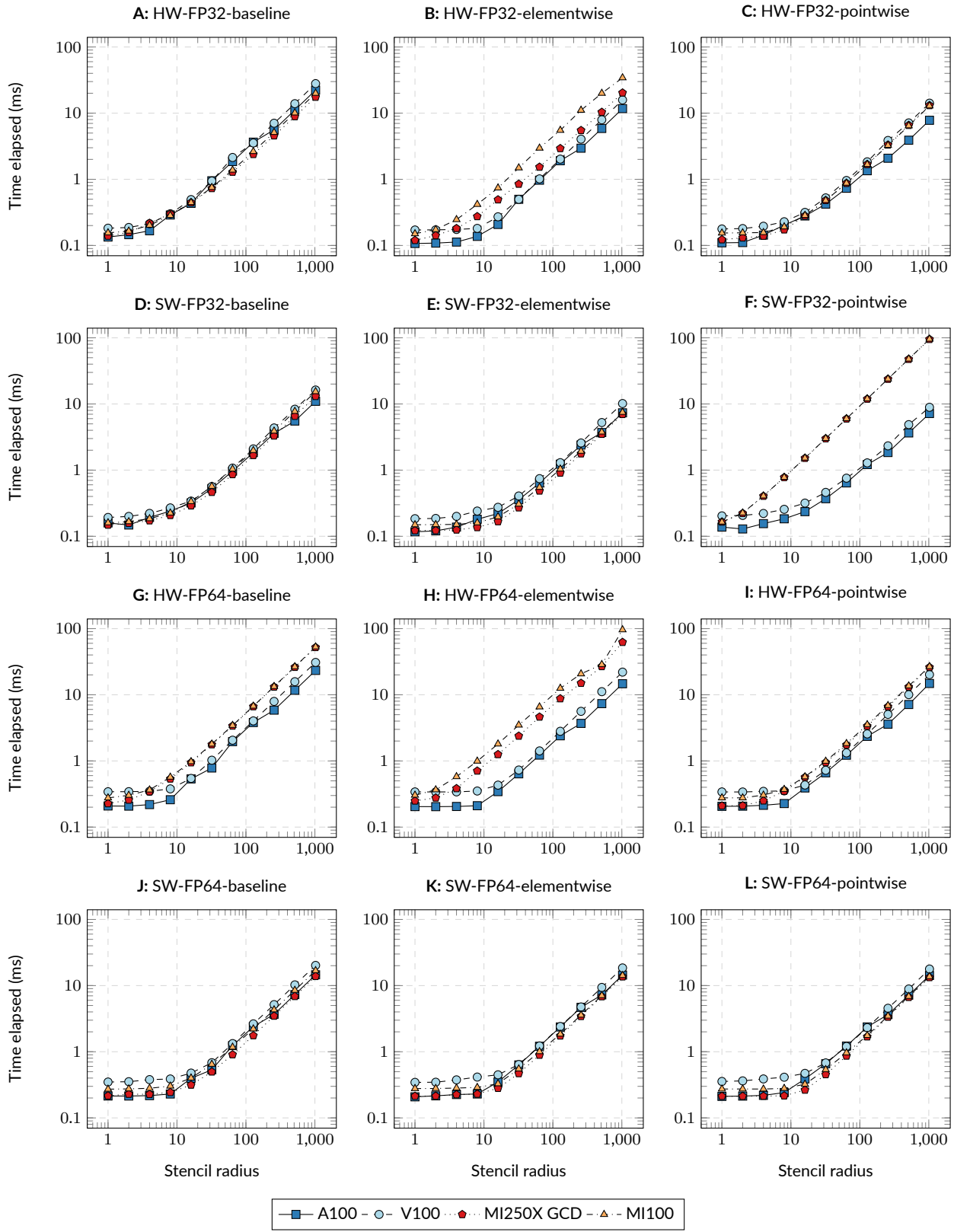

**FIGURE 9** A comparison of tuning strategies implemented for one-dimensional cross-correlations with CUDA and HIP. Lower time elapsed is better. Each subfigure is titled in the format [caching method]-[precision used]-[tuning strategy]. The hardware- and software-cached methods are denoted with *hw* and *sw*, respectively. The three evaluated tuning strategies are denoted as *baseline*, *elementwise*, and *pointwise*. We use *baseline* to denote the tuning strategy where each thread computes one output. In element-wise unrolling, shortened here to *elementwise*, each thread computes four outputs at neighboring locations. In stencil point -wise unrolling, shortened here to *pointwise*, each thread computes one output, but the multiply-accumulate loop over the stencil points has been unrolled.

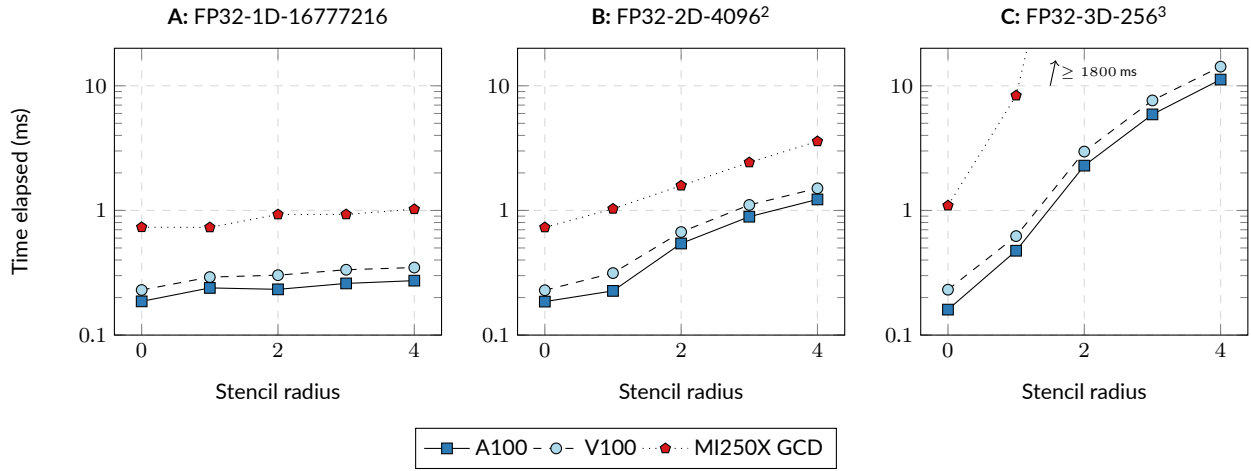

**FIGURE 10** Time per step for computing the diffusion equation in single precision with PyTorch. Lower is better.

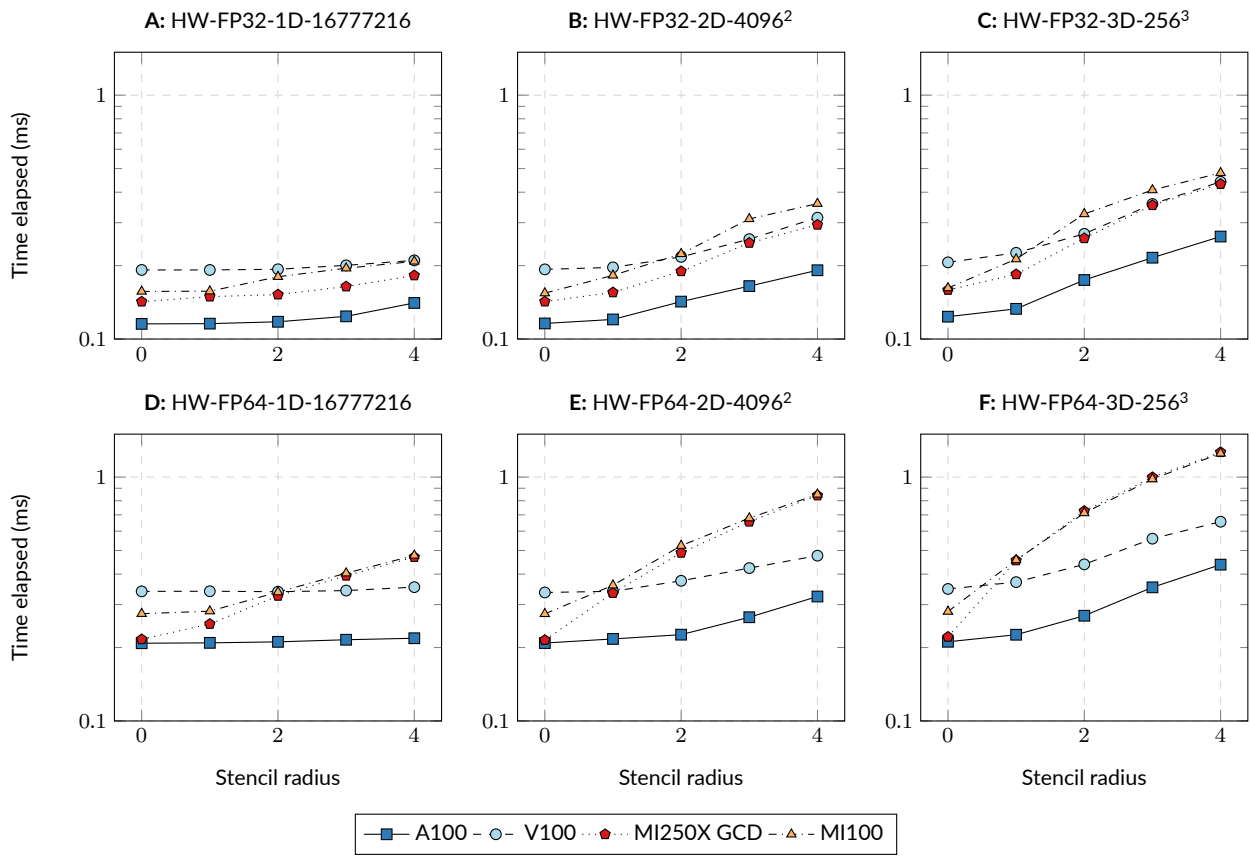

**FIGURE 11** Time elapsed per step in computing the diffusion equation with Astaroth in single and double precision. Lower is better.

## 6 | DISCUSSION

This article investigated the performance, energy efficiency, and tuning features of modern graphics processors designed for high-performance computing. The findings related to the differences in the hardware architectures are disseminated in Section 6.1. Section 6.2 focuses on the use of data tensors as the computational primitives for computing partial differential equations. Finally, the hardware- and software-cached strategies to fusing multiphysics kernels are discussed in Section 6.3.

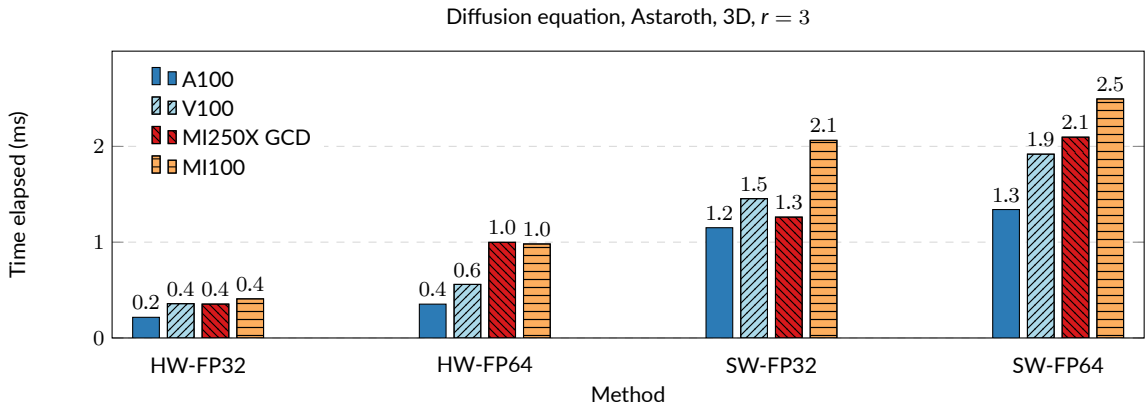

**FIGURE 12** Comparison of tuning strategies for computing the diffusion equation and implemented with Astaroth. Lower is better. The hardware-cached implementation provided the best performance on all devices.

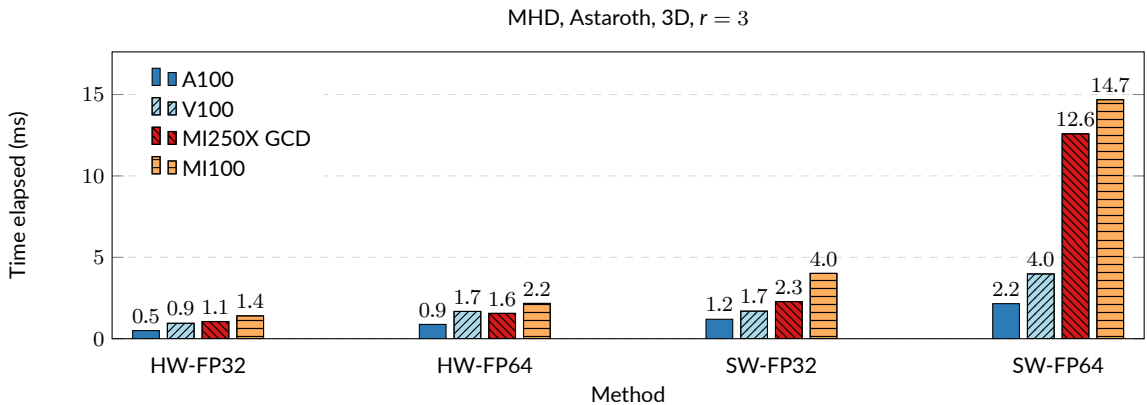

**FIGURE 13** Comparison of tuning strategies for computing the MHD equations with Astaroth. The plot shows the time elapsed for computing the final substep of explicit third-order Runge-Kutta integration. Lower is better.

**TABLE 3** Energy efficiency of the cross-correlation, diffusion equation, and magnetohydrodynamics cases as million element updates per second per watt. Higher is better. The calculations are based on the thermal design power (TDP) of the graphics processor published by the manufacturer. MI250X TDP has been halved to account for only one GCD in use. However, it should be noted, that achieving the energy efficiency reported here would require ideal scaling from one GCD to the full card.

| Case | Implementation | Dimensions | Precision | Radius | A100 | V100 | MI250X GCD | MI100 |
| --- | --- | --- | --- | --- | --- | --- | --- | --- |
| Cross-correlation | CUDA/HIP | 16777216 | FP32 | 1 | 391.3 | 326.4 | 500.8 | 374.1 |
|  |  |  | FP64 | 1024 | 3.0 | 3.1 | 4.5 | 4.1 |
| Diffusion equation | Astaroth | $256^3$ | FP32 | 1 | 315.4 | 247.8 | 325.2 | 263.0 |
|  |  |  | FP64 | 4 | 95.9 | 85.2 | 47.4 | 44.7 |
| MHD | Astaroth | $128^3$ | FP32 | 3 | 10.5 | 7.4 | 7.1 | 5.0 |
|  |  |  | FP64 | 3 | 6.0 | 4.2 | 4.8 | 3.2 |

## 6.1 | Hardware and software differences

As the first contribution, we measured the performance of the A100, MI100, MI250X, and V100 graphics processors in linear and nonlinear stencil computations in one- to three dimensions, placing special focus on the performance under varying cache loads. We applied common tuning strategies based on hardware- and software-managed caching. In all cases, better performance could be achieved with architecture-specific tuning. The effectiveness of the tuning strategies varied by device, the starkest differences arising between the devices of the two manufacturers.

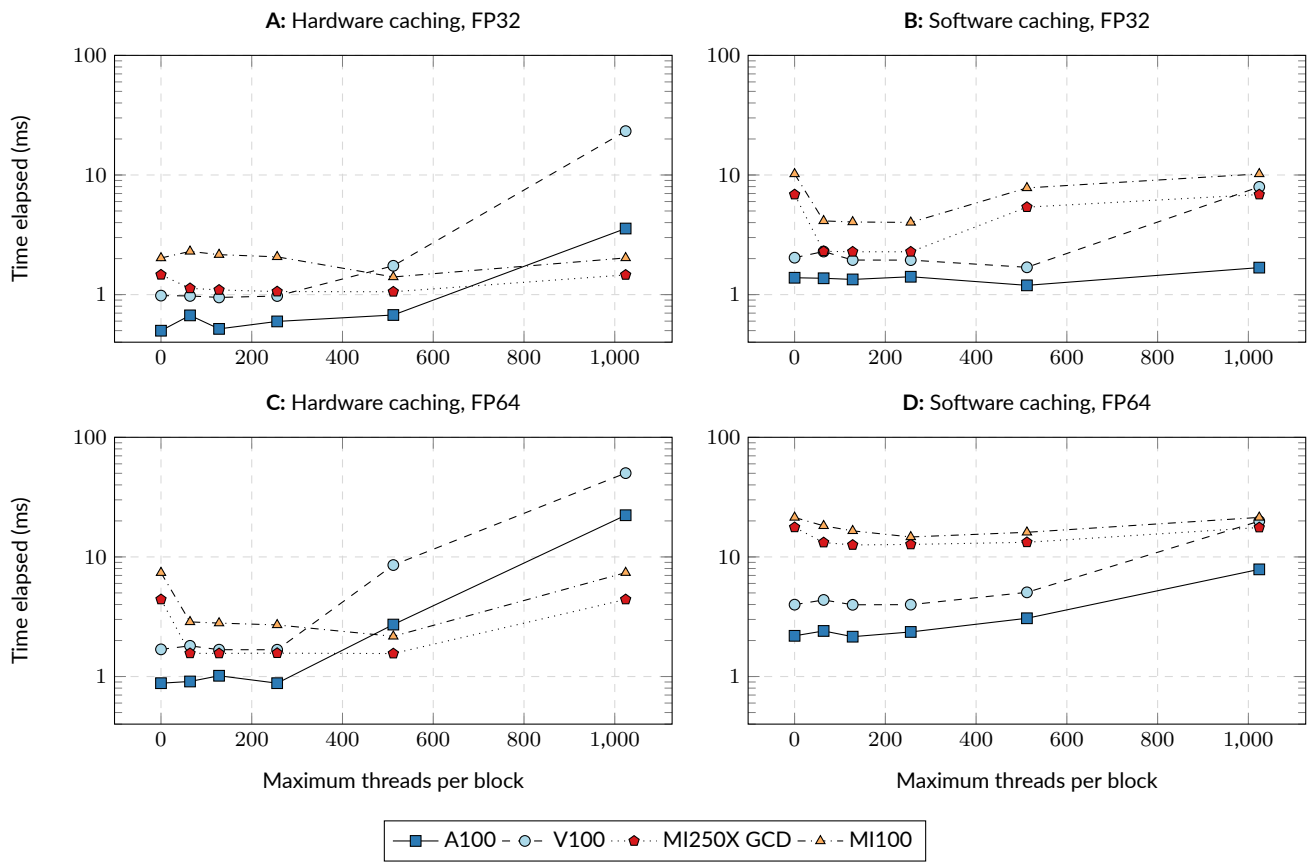

**FIGURE 14** Exploration of the `__launch_bounds__` tuning parameters with Astaroth in three-dimensional MHD simulations with $r = 3$. The results indicate the time elapsed for computing the final substep of explicit third-order Runge-Kutta integration. Lower is better. The performance with the default allocation without specifying `__launch_bounds__` is shown at position $0$ on the plot's *x* axis. In the ideal case, the best performance is achieved with the default register allocation without the need for manual tuning.

On the MI100 and MI250X, competitive performance could be achieved using software-managed memory for on-chip data reuse and manually tuning the per-thread register allocation. In comparison, the difference in performance between the initial, best, and worst implementations was less pronounced on the A100 and V100. Furthermore, we encountered severe performance pitfalls arising from seemingly benign code structures that affected only a part of the tested devices ‖. These findings emphasize the importance of performance validation on all target architectures when developing cross-platform software.

The aggregate on-chip bandwidth had a significant impact on the performance of the most cache-intensive benchmarks. In this aspect, the unified L1/shared memory architecture was a key difference between the devices of the two manufacturers. Historically, L1 bandwidth has been higher on GPU architectures where the L1 cache and shared memory reside on the same memory unit[29]. Nvidia states that the improved latency, bandwidth, and capacity of this design reduces the performance gap between hardware- and software-managed implementations[35]. The same effect has been independently noted for V100 GPUs in previous work[29]. We also observed, that due to higher L1 bandwidth, the performance gap between hardware- and software cached implementations was modest in cache-intensive computations on devices with unified L1/shared memory architecture.

On AMD CDNA 2 devices, the L1 cache is separate from the LDS and not on the compute unit[30], which hints at a lower bandwidth compared to the LDS. Our results also point at this direction, because the effective bandwidth of L1-bound code was lower compared to a kernel that utilized shared memory, as shown in Fig. 8. This highlights the importance of using shared memory to achieve competitive performance on architectures where the separation of L1/shared memory results in lower L1 bandwidth. The performance differences arising from differences in L1 capacities were more difficult to quantify based on tests selected for this work.

---

‖ See three-dimensional diffusion equation with PyTorch in Fig. 10C, stencil-point wise unrolling in Fig 9F, and conditional writes discussed in Section 5.4.

In the handtuned CUDA/HIP implementations, the performance of the A100, MI100, MI250X, and V100 graphics processors followed a ranking based on their L1, shared memory, and off-chip memory bandwidths depending on the cache-intensity of the benchmark. However, in a stark contrast with expectations, the MI100 and MI250X underperformed by several factors in cuDNN, MIOpen, and PyTorch benchmarks compared to the A100 and V100. Related work on convolution performance of modern AMD and Nvidia graphics processors is sparse, and we did not find conclusive evidence that would have either confirmed or refuted our findings. In one study, the authors reported that an A100 performed factors $1.5$ and $2.7$ faster in VGG-16 and ResNet-50 benchmarks, respectively, compared to an MI100[59]. These results are relatively close to our observations in Fig. 7, where we saw $2.6-3.69$ times speedup with a median of $3.13$. However, in another study, the MI100 was found to outperform the V100 in two-dimensional convolutions, including VGG and ResNet, when using single precision[60], whereas in our cuDNN/MIOpen benchmarks, the V100 gave consistently better performance. The AMD platform has been reported to be still maturing[61,60], which we suspect could contribute to these results. We conclude that in the tested cases, AMD graphics processors are capable of providing a factor $5.3 - 10.6$ better convolution performance on large input tensors and small batch sizes when the implementation is decoupled from the requirements of a machine learning library.

The capacity of the shared memory unit on current GPU architectures remains a limitation in applications that would benefit from extremely large caches. For example in MHD simulations, being able to store the full working set of a meaningfully large three-dimensional subdomain of the input tensor would improve on-chip reuse ratios by enabling the streaming cache optimization, where multiple output elements are computed with a single thread before evicting the data. Furthermore, if compute performance continues to increase at a faster rate relative to the performance of memory systems, more on-chip reuse opportunities have to be found to increase the operational intensity of kernels to match machine balance. Without increases to the capacity of shared memory, we expect this to pose a challenge, because a typical approach for increasing the operational intensity of an applications is to improve on-chip data reuse by increasing the amount of work per thread, which generally requires more on-chip memory for holding the intermediate results. In previous work, larger on-chip memories have also been stated to help in increasing the operational intensity in tensor processing applications[62].

## 6.2 | Data tensors as the computational primitives for PDEs

As the second contribution, we demonstrated an approach for reformulating the time integration of the diffusion and magnetohydrodynamic equations as a chain of tensor transformations. This approach provides an interesting view of the computations, because it enables the use of machine learning primitives for the direct numerical simulation of physical systems, transferring the burden of platform-specific tuning to highly-optimized software libraries that can leverage hardware specialized in tensor processing.

We implemented the chain of tensor transformations with PyTorch and Astaroth software libraries. While the PyTorch implementation was succesful as a proof of concept, further tuning, especially via kernel fusion and holding the working set near the chip throughout the computations, is needed for the implementations to be usable in production. The implementation used with Astaroth provided competitive performance suitable for production on all tested devices, and achieved performance within an order of magnitude of the theoretical upper bound that could be achieved with unlimited caches by utilizing the theoretical maximum off-chip bandwidth.

Utilizing tensor operations provided by machine learning libraries for the simulation of physical systems is under active research. However, the majority of work is focused on utilizing machine learning methods for creating physics-informed models, instead of using the data abstractions to map traditional numerical simulations to the hardware. In an approach similar to ours[63], the authors proposed a method for computing partial differential equations in simulations of computational fluid dynamics using TensorFlow, and applied their approach on tensor processing units. In another closely related work, the library `magnum.np` was used for finite-difference computations using PyTorch as the backend[64].

## 6.3 | Kernel fusion for multiphysics

As the final contribution, we proposed hardware- and software-cached strategies to fusing multiphysics kernels and performing the intermediate computations on or near the chip. We demonstrated a case, where competitive performance could be achieved without the use of shared memory. The difference in the performance of the HWC and SWC implementations was largely explained by the difference in instruction counts. The SWC implementation required additional instructions for index calculations when loading the data to shared memory. The keys to improving performance in both implementations was to reduce instruction overheads by unrolling computations, using local memory for caching intermediate results, and reordering instructions to improve instruction-level parallelism to cover for low occupancy due to heavy register use. Additionally, automatic tuning enabled finding the best-performing domain decomposition for the target architecture.

Performance optimization of stencil computations has been studied extensively in previous works[5,65,66,67,68,69]. The concept of achieving better performance on graphics processors by trading occupancy for improved register caching via unrolling and instruction-level parallelism is an optimization technique popularized in the early 2010s [31]. As demonstrated with our implementation of MHD, this technique is also effective for optimizing multiphysics kernels.

# 7 | CONCLUSION

We investigated the performance of stencil computations on graphics processors. Our results highlighted the necessity of performance evaluation and tuning on the target platforms when developing cross-platform software. Furthermore, we highlighted the use of data abstractions native to graphics processors to guide performance tuning and reduce burden of optimization. Finally, we proposed a strategy for fusing kernels in multiphysics simulations. Our implementation demonstrated, that with a set of platform-independent techniques focused on reducing instruction counts, improving instruction-level parallelism, and utilizing local memory heavily for intermediate computations, competitive performance can be achieved without using shared memory.

**ACKNOWLEDGEMENTS**

This project has received funding from Academy of Finland, ReSoLVE Centre of Excellence (grant number 307411), the European Research Council (ERC) under the European Union's Horizon 2020 research and innovation programme (Project UniSDyn, grant agreement n:o 818665), and KAUTE foundation (grant number 20240173). We acknowledge the computational resources provided by CSC — IT Center for Science and the Aalto Science-IT project.

**DISCLOSURE OF THE USE OF AI TOOLS**

Large language models (LLMs) ChatGPT 3.5 [70] and Microsoft Copilot [71] were used for assistance in the preparation of this article. All responses were treated as an unreliable third-party source and cross-checked with existing literature. The article does not contain verbatim responses of LLMs and the authors take full responsibility that the text is original and reflects their best knowledge. AI tools were used in the following ways.

- In limited instances, as assistance to language editing by providing grammar, sentence structure, and wording suggestions for expressing the authors' original idea.
- As assistance for studying for the work by providing additional clarifications and summaries about well-documented topics.
- As assistance for finding additional arguments against the authors' views.
- As assistance for finding suggestions for additional topics to discuss.

**CONFLICT OF INTEREST**

The authors declare no conflicts of interest.

20 | Pekkilä et al.

# APPENDIX

## A MAGNETOHYDRODYNAMIC EQUATIONS

The equations for the magnetohydrodynamic setup are listed in Eq. A1, A2, A3, and A4. The symbols are listed in Table A1. Readers should take note that the notation here follows the conventions of physics literature and differs from the conventions used throughout the rest of the paper.

**TABLE A1** A list of symbols used in Eqs. A1, A2, A3, and A4.

| Description | Symbol |
| --- | --- |
| Logarithmic density | $\ln \rho$ |
| Velocity | **u** |
| Specific entropy | $s$ |
| Magnetic vector potential | **A** |
| Laplace operator | $\nabla^2$ |
| Curl operator | $\nabla \times$ |
| Tensor contraction operator | $\otimes$ |
| Advective derivative | $D/Dt = \partial/\partial t + \mathbf{u} \cdot \nabla$ |
| Magnetic field | $\mathbf{B} = \nabla \times \mathbf{A}$ |
| Magnetic diffusivity | $\eta$ |
| Magnetic vacuum permeability | $\mu_0$ |
| Electric current density | $\mathbf{j} = \mu_0^{-1} \nabla \times \mathbf{B}$ |
| Traceless rate-of-shear tensor | **S** |
| Specific heat capacity at constant pressure | $c_p$ |
| Specific heat capacity at constant volume | $c_v$ |
| Kinematic viscosity | $\nu$ |
| Bulk viscosity | $\zeta$ |
| Adiabatic speed of sound | $c_s$ |
| Adiabatic index | $\gamma$ |
| Explicit heating term | $\mathcal{H}$ |
| Explicit cooling term | $\mathcal{C}$ |
| Radiative thermal conductivity | $K$ |
| Temperature | $T$ |

**TABLE B2** Configuration used for running and verifying the benchmarks. Column $\Delta s_i$ denotes grid spacing in dimension $i$. Units in the last place is abbreviated as ULP. We use $\epsilon$ to denote the IEEE 754 (2008) interval machine epsilon for the same precision the calculations are performed in. If necessary, we denote the machine epsilon for single- and double-precision numbers with a subscripts $\epsilon_{32}$ and $\epsilon_{64}$, respectively. Superscript $\epsilon^{(C)}$ is used to denote the machine epsilon specified by the ISO C standard and defined as either `FLT_EPSILON` or `DBL_EPSILON` in the C standard library.

| Library | Case | Init. range (verification/benchmarks) | $\Delta s_i$ | $\Delta_t$ | Acceptable error |
| --- | --- | --- | --- | --- | --- |
| Astaroth | Diffusion equation | $[0, 1] / (-10^{-5}, 10^{-5}]$ | $2\pi/n_i$ | $\epsilon_{32}^{(C)}$ | Rel. error < 5 ULP or abs. error < $\epsilon^{(C)} \min_i \mathbf{f}_i$ |
|  | MHD | $[0, 1] / (-10^{-5}, 10^{-5}]$ | $2\pi$ | $\epsilon_{32}^{(C)}$ | Rel. error < 5 ULP or abs. error < $\epsilon^{(C)} \min_i \mathbf{f}_i$ |
| PyTorch | Diffusion equation | $[-1, 1)$ | $2\pi/n_i$ | $10^{-3} \min_i(\Delta s_i)$ | $\|a - b\| \leq (5\epsilon + 5\epsilon\|b\|)$ |
|  | MHD | $[-10^{-3}, 10^{-3})$ | $2\pi/n_i$ | $10^{-1} \min_i(\Delta s_i)$ | $\|a - b\| \leq (100\epsilon + 100\epsilon\|b\|)$ |

**TABLE C3** Relative performance of PyTorch to cuDNN/MIOpen in one-dimensional cross-correlations. PyTorch is faster at values < 1.

| radius | A100 | V100 | MI250X GCD |
| --- | --- | --- | --- |
| 1 | 1.07 | 1.04 | 1.16 |
| 2 | 0.90 | 0.98 | 1.13 |
| 4 | 0.86 | 0.90 | 1.08 |

The implementation follows closely the one presented in [6]. For more details on the computational aspects, we refer the reader to [44].

$$\frac{D \ln \rho}{Dt} = -\nabla \cdot \mathbf{u} \tag{A1}$$

$$\frac{D\mathbf{u}}{Dt} = -c_s^2 \nabla \left( \frac{s}{c_p} + \ln \rho \right) + \frac{\mathbf{j} \times \mathbf{B}}{\rho} + \nu \left[ \nabla^2 \mathbf{u} + \frac{1}{3} \nabla(\nabla \cdot \mathbf{u}) + 2\mathbf{S} \cdot \nabla \ln \rho \right] + \zeta \nabla(\nabla \cdot \mathbf{u}) \tag{A2}$$

$$\rho T \frac{Ds}{Dt} = \mathcal{H} - \mathcal{C} + \nabla \cdot (K \nabla T) + \eta \mu_0 \mathbf{j}^2 + 2\rho \nu \mathbf{S} \otimes \mathbf{S} + \zeta \rho (\nabla \cdot \mathbf{u})^2 \tag{A3}$$

$$\frac{\partial \mathbf{A}}{\partial t} = \mathbf{u} \times \mathbf{B} + \eta \nabla^2 \mathbf{A} . \tag{A4}$$

## B  ADDITIONAL SETUP INFORMATION
## C  ADDITIONAL BENCHMARKS

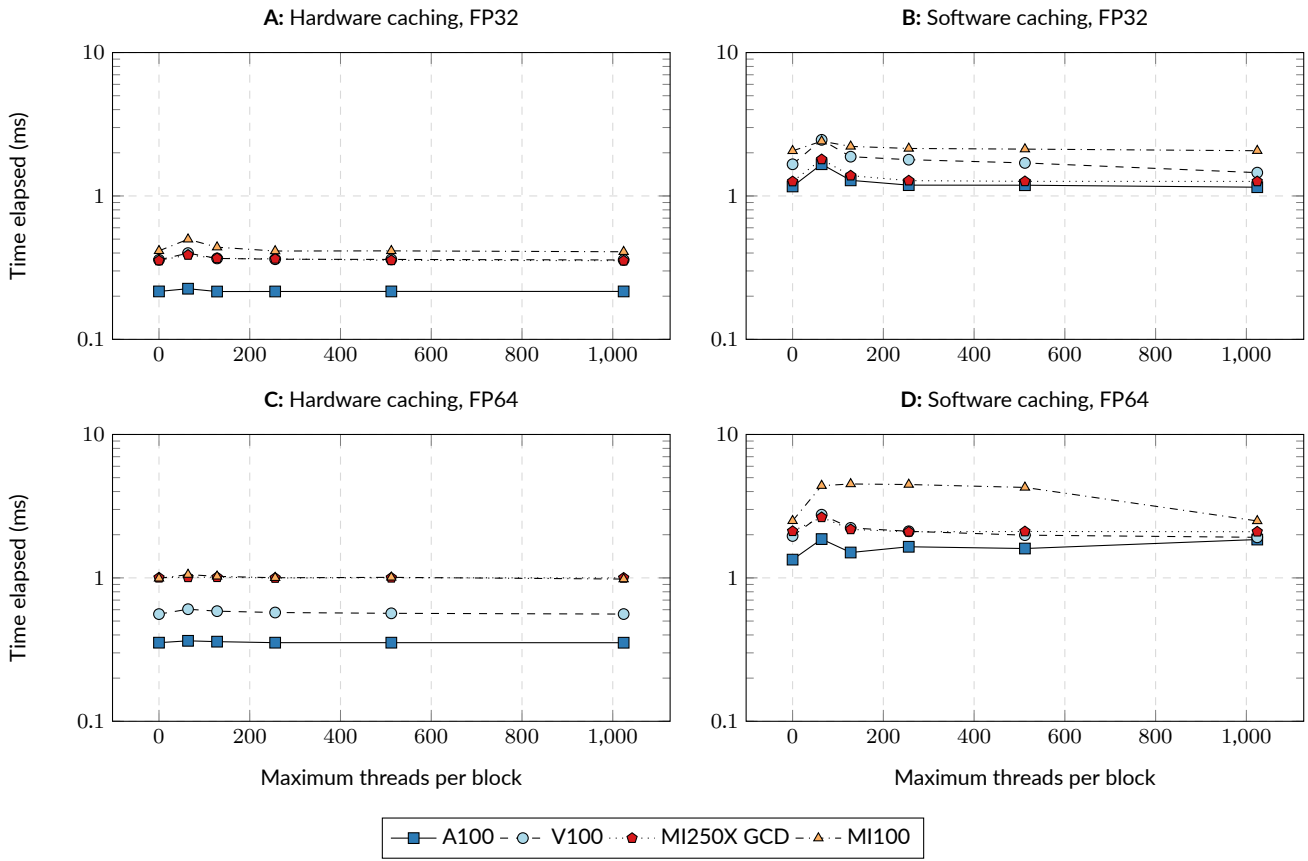

**FIGURE C1** Exploration of the tuning parameters for computing the diffusion equation with Astaroth. Lower time elapsed is better. Register allocation was controlled by defining the maximum thread block size with the `__launch_bounds__` kernel qualifier. Performance at the default register allocation without `__launch_bounds__` is plotted at position $x = 0$.